# Increased and Varied Radiation during the Sun's Encounters with Cold Clouds in the last 10 million years

**Authors:** Merav Opher[1,*], Joe Giacalone[2], Abraham Loeb[3], Evan P. Economo[4,5], Alan Cummings[6], Jennifer Middleton[7], Catherine Zucker[8], Jesse A. Miller[1], Anna Nica[1], Maria Hatzaki[9]

[1,*]Astronomy Department, Boston University, Boston, MA 02215, USA

[2]Lunar and Planetary Laboratory, University of Arizona, Tucson, AZ, USA

[3]Astronomy Department, Harvard University, 60 Garden Street, Cambridge, MA, USA

[4]Department of Entomology, University of Maryland, College Park, MD, USA

[5]Biodiversity and Biocomplexity Unit, Okinawa Institute of Science and Technology Graduate University, Onna, Okinawa, Japan

[6]California institute of Technology, USA

[7]Lamont-Doherty Earth Observatory, Columbia University, Palisades, NY, USA

[8]Smithsonian Astrophysical Observatory, Cambridge, MA, USA

[9]National and Kapodistrian University of Athens, Athens, Greece

*Correspondence to: Merav Opher, mopher@bu.edu

**Abstract:**

Recent research raises the possibility that 3 and 7 million years ago, the Sun encountered massive clouds that shrank the heliosphere —the solar cocoon protecting our solar system— exposing Earth to its interstellar environment, in agreement with geological evidence from $^{60}$Fe and $^{244}$Pu isotopes. Here we show that during such encounters Earth was exposed to increased radiation in the form of high-energy particles. During periods of Earth's immersion in the heliosphere, it received particle radiation that we name Heliospheric Energetic Particles (HEPs). The intensity of < 10 MeV protons was at least an order of magnitude more intense than today's most extreme solar energetic particle (SEP) events. SEPs today last minutes to hours, but HEP exposure then lasted for extensive periods of several months, making it a prolonged external driver. During Earth's excursion outside the heliosphere, it was exposed to a galactic cosmic ray radiation with the intensity of < 1 GeV protons at least an order of magnitude more intense than today. Therefore, the space surrounding Earth was permeated by a variable high-energy radiation. We discuss the implications for Earth's climate and biodiversity.

**Introduction:**

Stars are not stationary but rather move through their surrounding media. Stars also have winds. A star's motion, coupled with its wind, creates a protective cocoon known as an "astrosphere." Referred to as the heliosphere, the Sun's astrosphere protects the Earth and other planets in our solar system. The heliosphere shields the stellar system from low energy galactic cosmic rays (GCRs) and interstellar dust. Indeed, the current heliosphere safeguards Earth from 70% of GCRs with energy > 70 MeV/nuc (see Figure 1 of ref 1). During its lifetime, however, the Sun has traveled through various regions and the different environments it has encountered must have affected the size and nature of the heliosphere. Space physicists mainly focus on understanding today's heliospheric processes and their effects on radiation that would matter for astronauts, or on understanding other space weather effects in today's solar system but have not focused on reconstructing historical heliospheric dynamics and downstream consequences.

There have been suggestions that the Sun will encounter massive and dense clouds every billion years[2,3], and be affected by supernova explosions[4-6]. The Sun's movement and exposure to different interstellar environments might profoundly affect the heliosphere and Earth.

There have been attempts to track the location of the Sun and the associated impact of its position on the cosmic ray flux to Earth[7,8]. Some works have explored potential increases of radiation as the Sun passes through galactic arms every ~ 140 Myrs[9]. However, without mapping the Sun's trajectory and constraining the precise locations of structures with which the Sun could have collided, these previous works were suggestions based on probabilities.

Recently several works mapped the trajectory of the Sun in the last 10 million years and showed that there is a possibility that the Sun crossed the path of massive cold clouds 2-3 and 6-7 million years ago[10,11]. This work incorporated major advances in astronomical data such as the *Gaia* mission that precisely records the motion of stars and their interstellar environments[12].

With state-of-the-art magnetohydrodynamic (MHD) simulations[13], these works showed that the heliosphere collapsed to sub-AU scales during the Sun's passage through these dense cold clouds. This collapse left Earth exposed to the ISM and the contents of the cloud. The timing of these encounters agrees with radioactive $^{60}$Fe isotopes and $^{244}$Pu found in deep sea sediments[14,15] and ferromanganese crusts, in Antarctic ice[16], in Galactic cosmic rays[17], and in lunar samples[18] that indicate an increase in abundance around 2–3 and 6-7 million years ago.

As result of encounters with cold clouds, Earth would have been exposed to large amounts of neutral hydrogen with densities above 1,000 cm$^{-3}$, which may have altered the chemistry and climate of the Earth's atmosphere[19,3]. Past work predicted that such high densities would deplete the ozone in the mid-atmosphere and eventually cool the Earth[3,19]. This predicted cooling coincides with the intervals of global cooling between 2-3 and 6-7 Ma inferred from oxygen isotopes measured in foraminifera on the sea floor[20-21], for which the driving mechanisms remains a subject of debate. These earlier[3,19] estimates need to be revisited with modern climate models. Recently, we found that the effect on climate, while drastic, is more complex[22], with noctilucent cloud formation significantly enhanced and global depletion of ozone in the mesosphere.

We note that another scenario proposed to explain the excess in abundance of $^{60}$Fe and $^{244}$Pu is that it is due to ejecta from a nearby supernova. The supernova ejecta scenario requires that two supernovae exploded within ~100pc of the Sun at 3Mya and 7Mya[23]. The responsible supernovae could not have exploded closer than 8pc, which is the so called "kill radius" that would initiate a mass extinction[24]. This scenario also requires that $^{60}$Fe condense into dust grains in the interstellar medium so that it could penetrate the heliosphere and deposit on Earth without compressing the heliosphere to <1 au[25-27,15,4]. These models include further assumptions involving the delivery of dust to the Solar System and its entry into the heliosphere that are still open areas of research. According to recent modelling efforts, the Local Bubble was driven by a series of ~15 supernovae that started exploding circa 14 Mya[28]. This presents an interesting challenge to the supernova scenario as to why only two $^{60}$Fe pulses have been detected in the last 10 Myr whereas Local Bubble modelling implies a rate of approximately 1 per Myr.

In the cold cloud scenario supernova explosions deliver $^{60}$Fe to the interstellar medium in form of dust that gets incorporated into massive clouds existent in the interstellar medium. The difference between this scenario and other competing scenarios is that the supernova explosions don't have to coincide with 2-3 Mya or 6-7 Mya. As Earth goes through a massive cloud enriched with $^{60}$Fe, the flux of terrestrial $^{60}$Fe will increase.

It is interesting to note that there works that suggest that encounters with dense interstellar clouds could temporarily shut off the solar wind due to high accretion rates[29-31,2]. For instance, an encounter of a cloud with neutral density >330 cm$^{-3}$ with a relative speed between the cloud and the Sun of 20 km/s, was predicted to halt the solar wind[30]. These suggestions should be revisited with modern models, as well with Alfven driven winds since they did not consider the presence of the heliosphere and effects of charge exchange or photoionization between neutral hydrogen atoms and ions.

Here we explore the implications that a collapsed heliosphere would have for the high-energy charged particle radiation in the near-Earth space. We focus on the period of 2-3 Mya but the results as discussed below should be valid for other encounters of the Sun and heliosphere with cold clouds. The next section will describe the models used. The following section describes the varied and increased particle radiation that Earth would be subjected to during its orbit as it crosses the cold cloud. In the final section, we discuss its implications.

**Methods**

There are three different models used in this paper.

1. MHD model, just to model the heliosphere in a dense cold cloud. This only models the heliosphere and not energetic particles.

2. The hybrid simulation, used only to model the acceleration of the solar wind locally at the termination shock at one place. This is an extremely local simulation. Its purpose is only to do the proper kinetic physics and the formation of the suprathermal tail in the heated solar wind distribution at the shock. This simulation allows us to normalize our results for the particle acceleration model.

3. Particle acceleration model: this is the solution to the Parker equation. We use the results of model 2 as a normalization.

**Description of the 3D MHD model**:

Here we just briefly summarize the model used; for more details see ref 10. We used a state-of-the-art 3D MHD model that considers a single ionized component and four neutral components, although for the run we used only the ISM component which is orders of magnitude more abundant than the heliosheath and supersonic components. We used inner boundary conditions for solar wind at 0.1 au (or 21.5 solar radii). The parameters adopted for the solar wind were based on the well-benchmarked Alfven-driven solution.

The grid was highly resolved at $1.07 \times 10^{-3}$ au near 0.1 au and $4.6 \times 10^{-3}$ au in the region of interest, including the tail. 2-3Mya the Solar System may have passed through the Local Ribbon of Cold Clouds (LRCC) in the constellation of Lynx[10] – we named that portion the Local Lynx of Cold Clouds (LxCCs). Local Leo Cold Cloud (LLCC)[32] is among the largest and most studied clouds of the LRCC. We don't know the origin of these clouds although LRCC has a very placid and smooth velocity field, so for the ISM outside the heliosphere, we adopted the characteristics of the LLCC[32], namely, $n_H = 3,000$ cm$^{-3}$ and T = 20 K. We included a negligible ionized component ($n_i = 0.01$ cm$^{-3}$ and T = 1 K) and ignored the interstellar magnetic field as its pressure is negligible compared to the ram pressure of the cold cloud. Similarly to the LLCC[32], we assume that the LxCC to have a very small H$_2$ density of ≈1.5cm$^{-3}$ compared to its atomic hydrogen density. For numerical stability we initiated with ISM filling out the whole domain while the solar wind is initiated from the inner boundary blowing into the ISM conditions. The run was performed for 44 years.

We adopted the relative speed between the Sun and LxCCs (ΔU, ΔV, ΔW) = (−13.58, −1.40, 3.70) km s$^{-1}$ or a velocity of 14.1 km s$^{-1}$. We rotated the system so that the flow is in the z–x plane with the ISM approaching from the -x direction. The neutral H from the cold cloud impinged on the heliosphere with speeds of $U_x$ = 14.1 km s$^{-1}$, $U_y$ = 0 km s$^{-1}$ and $U_z$ = 1.1 km s$^{-1}$. The numerical model includes charge exchange between the neutrals and ions, as well as the Sun's gravity. Neutral H atoms are included through a multi-fluid description that is appropriate for the high densities. Two fluids are used, one between the pristine ISM and the bow shock that forms ahead of the heliosphere and one that captures the heated and decelerated population between the bow shock and the heliopause (HP). The heliosphere shrinks to 0.22 ± 0.01 au, which is well within the Earth's orbit, thus exposing the Earth (and all the other Solar System planets for most of their trajectories) to the ISM, which has neutral densities of 3,000 cm$^{-3}$. The heliosphere has a long tail that extends for several au's so for part of its trajectory Earth, is deep in the solar wind subjected to solar wind particles accelerated by the closed-in Termination Shock that we named "heliospheric energetic particles" or HEPs[85]. HEP refers to particles accelerated in the outer heliosphere, to distinguish it from SEP, or solar-energetic particle, which is of solar origin. There are other factors that make it a more-appropriate for this paper as well, most notably that the source of these particles for the case of a termination shock at 0.118AU is the solar wind, and not interstellar pickup ions, as is the case for today's termination shock. Here we are interested in evaluating the radiation.

**Description of the Hybrid simulation:** The details of the model can be found in ref 33. In this approach, the protons are treated kinetically and the electrons as a fluid. A shock is formed in this model by forcing plasma into a rigid wall at one end of the simulation domain. The simulation domain was taken to be $10^4$ proton inertial lengths, which is about 0.0007 au for this case. The simulation was run long enough that the downstream distribution of protons achieved a steady state. This model has been applied to the physics of shock waves for over 40 years[86] and has successfully reproduced in situ spacecraft observations of ion distributions and fields at shocks in space, such as Earth's bow shock.

**Model which is a solution to the well-known Parker transport equation[34]:**

This model includes most of the major transport effects of cosmic rays including energy change, diffusion, and advection with the solar wind. In this calculation, we used a model that is very similar to that described by ref. 35 but uses a stochastic integration technique, as described by ref 36-37. Our model is spherically symmetric in space and the only spatial coordinate is heliocentric distance. The Termination Shock is represented by a change in the solar wind speed from a value of 418.7 km/s inside the Termination Shock radial location, to a value 115.4 km/s beyond it, in the downstream region. The cosmic rays are assumed to have a diffusion coefficient of the form $\kappa(E)=\kappa_0(E/E_0)$ where $\kappa_0=5\times10^{17} cm^2/s$, and $E_0=100$ keV. It is assumed to be independent of radial distance, except that it is taken to be 10% smaller downstream of the Termination Shock to mimic the effect of increased magnetic-field turbulence in the heliosheath. There are two spatial boundaries imposed, one at 2 solar radii, and the other at 1.25 times the distance of the Termination Shock, where absorbing boundary conditions are imposed. A continuous source at the shock is assumed with an energy of 8 keV.

It is worth noting that the mean-free path and gyroradii of the MeV particles from the Sun are much less that 1 **au**. For example, ref. 38 found a mean-free path of particles at a shock 0.35au from the Sun to be less than 0.1au at 1MeV, and ref. 39 finds something similar. The gyroradius is much smaller than these values. Because the mean-free path is so small that the Parker transport equation is valid. The Parker equation only needs the distribution to be isotropic to be valid, and the smaller the mean-free path, the more scattering, and the more likely the distribution is isotropic.

**Results**

Our model results suggest that the encounter of the heliosphere with the Local Lynx of Cold Clouds (LxCC), 2-3 million years ago shrank the heliosphere. The 3D MHD model used for this work is described in Methods, and ref. 10.

Figure 1 shows the contours of speed in the meridional and equatorial plane. One can see that the solar wind termination shock (the transition from red to blue/green) moves to distances as close as 0.118 au. The heliopause moves in as close as 0.22 au at the nose. The heliosphere has a long heliotail to the right – a zoom **out** is shown in Figure 1b. As the Earth travels around the Sun (red circle – Figure 1b), it dips in and out of the heliosphere immersed in a dynamic solar wind.

During periods when Earth is immersed in the collapsed heliosphere, it experiences radiation from particles accelerated at the Termination Shock. In today's heliosphere, the Termination Shock is known to accelerate particles up to MeV energies[40]. In the collapsed heliosphere the Termination Shock is much closer to the Sun (Figure 1a; c), being a much stronger shock than in today's heliosphere. The Termination Shock has a compression ratio of 3.6 (Figure 2a) in the nose and 3.2 in the flanks. The shock in this case is not mediated by the presence of the pick-up ions as in today's heliosphere that weaken the shock to a ratio of ~ 2 (ref. 40). The lack of pick-up ions occurs because with an ISM density ~3000 cm$^{-3}$ the mean free path due to charge exchange decreases to ~ 0.01-0.1 au and the neutrals do not come far into the heliosphere before being ionized. Consequently, we only considered charge exchange outside the heliopause in the current model (see Figure 6c of ref. 10).

In the collapsed scenario, the Termination Shock occurs at such close distances to the sun that the solar magnetic field is mainly radial, as opposed to the nearly azimuthal magnetic field where today's Termination Shock is located. This is because the radial component of the solar magnetic field falls with distance as $r^{-2}$, while the azimuthal component falls as $r^{-1}$ due to the rotation of the Sun. As a result, the unit normal to the shock in this case is nearly parallel to the field, which is known as a quasi-parallel shock, as opposed to the quasi-perpendicular Termination Shock of today.

To simulate the acceleration of particles to high energies at the Termination Shock of 2-3 million years ago, we used a combination of a self-consistent plasma simulation of a quasi-parallel shock, to treat the physics of heating and formation of suprathermal tails in the heated distribution at the shock, and a separate model, based on a solution to the cosmic-ray transport equation, to determine the distribution of much higher energy particles that are accelerated by the Termination Shock.

We used the well-known hybrid simulation[41,42] for the first of these calculations, specifically using the same model as in ref 43. For details see Methods.

We considered a one-dimensional simulation with parameters consistent with the solar wind at a distance of 0.118 au and based on the MHD simulations discussed above. The resulting shock-frame parameters are the Alfven Mach number, $M_A=3$, total (electron plus proton) plasma beta, $\beta_e+\beta_p= 0.4$, Alfven speed, $v_A=140$ km/s, and shock-normal angle, $\theta_{Bn}=0$. Shown in Figure 3(a) are snapshots at the end of the simulation of the magnetic field components (top), solar wind radial velocity (middle), and solar wind density (bottom). Note that in this one-dimensional calculation, the radial component is constant in order to preserve the divergence-free condition of the magnetic field.

The simulated differential intensity of protons is shown as a solid black curve in Figure 3(b). The initial solar wind distribution is shown as a dashed black curve. Note that there is considerable broadening of the distribution across the shock. In addition, the distribution is non-Maxwellian and has a suprathermal tail which begins at about 8 keV where there is a slight inflection. This is common in these simulations and has been well discussed in the literature[44,45]. At energies beyond about 10-20 keV, the distribution falls off sharply, which is largely because of the reduced spatial dimension of the simulation. The maximum energy achievable is determined by how easily the particles escape the system and depends on the efficiency of pitch-angle scattering, and the resulting parallel mean-free path of scattering, due to their interaction with magnetic fluctuations.

In order to determine the acceleration of very high-energy particles, whose mean-free paths and gyroradii are far larger than the computationally restricted hybrid simulation, we used a separate model which is a solution to the well-known Parker transport equation[34]. This general approach was considered by ref 46 to model anomalous cosmic rays at today's Termination Shock. For details see Methods. A steady-state solution to the Parker equation is obtained since the Termination Shock is assumed to be steady and no other time dependence is introduced. The red histogram in Figure 3(b) shows the resulting spectrum at the TS from this calculation. The Parker equation does not address the question of injection at low energies; thus, we have arbitrarily placed the red histogram at about the place where the high-energy tail in the hybrid simulation forms. This is the inflection point in the downstream distribution that we noted above. Thus, we assume that the hybrid simulation provides the physics of injection, while the Parker-transport model provides the physics of particle acceleration to high energies. Note that the spectrum falls off sharply at very high energies, where there is a spectral break. This spectral cut-off energy is determined by the value of the diffusion coefficient at the highest energies and the spatial size of the system, and is above 30 MeV for the parameters used in this study

We also note that the acceleration of solar wind to high energies is not merely just theoretical, or the result of a model calculation, but is also well documented by in situ spacecraft observations. Earth's bow shock is one particularly relevant example[47-49], as are interplanetary shocks[50,51].

Also shown in Figure 3(b) are data taken from the ACE/EPAM instrument for a large solar-energetic particle event (SEP) that occurred on DOY 302 2003. The data, covering energies from a several tens of keV to a few MeV, were taken immediately after that passage of a large - driven coronal mass ejection (CME) shock associated with this event. They are the same data as those presented in Figure 2 of ref 35. This event is among one of the largest SEP events observed in the space-era. As one can see by the red histogram, our prediction for the intensity of heliospheric energetic particles some 2-3 million years ago at the TS is ~ 2 orders of magnitude more intense than the 2003 Oct 29 SEP event. We refer to these as "heliospheric energetic particles" or HEPs to distinguish them from Solar Energetic Particles (SEPs) that are produced near the Sun and Anomalous Cosmic Rays (ACRs) that are accelerated at today's Termination Shock.

The simulated spectra shown in Figure 3(b) are the prediction for 0.118 au. Obtaining the value at 1 au is not simple owing to the nature of the morphology of the heliosphere during this extreme time period. The distribution of particles in the shocked solar wind should be approximately uniform along streamlines of the flow. Thus, within the heliosheath, we expect that the intensity of high-energy particles will be similar to that at the Termination Shock. Those particles which cross the heliopause and into the interstellar medium, in contrast, will rapidly escape along interstellar magnetic field lines. Although the heliosheath is highly structured, as evidenced by Figure 1(b), the intensity of particles accelerated at the TS should be similar to that even far from it, deep into the heliosheath. Thus, our predicted spectra shown in Figure 3(b) are about what Earth would have experienced each time it was immersed within the heliosheath, which could have been for a few months out of each year.

During periods when Earth will be exposed to the interstellar medium it will be subjected to cosmic rays pervading the Galaxy. Figure 4 shows in black the cosmic-ray radiation that Earth will experience once outside the heliosphere, adapted from Figure 3 of ref 1. For periods when Earth is immersed in the heliosphere the Earth will be subjected to the spectrum shielded by the heliosphere. In addition to the HEPs from the TS, Figure 4 shows today's GCR measurements at 1 au (BESS/IMP8) and well as HEPs. One can see that today's heliosphere shields the ISM GCR spectrum by up to many orders of magnitude at energies less than 1GeV. Here we assume that the GCR spectrum in the cold cloud will be the same as in today's interstellar medium. We make a first order assumption that the absorption of GCRs by the LxCC is negligible: if the size of the LxCC is between 200 **au** and 1 pc, a 1 GeV proton will have a mean free path of 2,700 pc crossing through the cloud, and thus collisions with atomic hydrogen are unlikely. We assume here that the filtration of GCR in the collapsed heliosphere will be similar.

The issue as to what causes the filtration of low energy GCRs is one of the main questions in today's heliospheric science[13]. Works such as ref 52 examine which radial gradients are needed to explain the filtration of GCR as measured by the Voyager spacecraft. For today's heliosheath they obtained radial gradients of 1.5%-1.8% per au.

This work shows that the radiation that Earth would experience from changes in the heliosphere would be a combination of increased GCRs for energies < 1 GeV for a large portion of its orbit and increased energetic particles (HEPs) when immersed in the heliosheath. One can see from Figures 3 and 4 that the energetic particles at 10 MeV at 1 au would be some 7-8 orders of magnitude more intense than GCRs of the same energy. The red curve is ~2 orders of magnitude more intense than the 2003 Oct 29 SEP event which was one of the largest recorded. There is no doubt that at 10 MeV, large SEP events are far more intense than GCRs. The penetration of MeV particles to Earth depends on the Earth's magnetosphere as we discuss below.

**Conclusions and Discussion**

It is interesting to explore if such radiation (HEPs and GCRs) can be detected in historical records because intense solar energetic proton events (and GCR) can create cosmogenic isotopes like $^{10}$Be (that has a half-life of 1.4 x 10$^6$ years) and $^{14}$C (that has a half-life of 5730 years)[53-55]. To investigate this, we must consider the likely duration of such particle radiation. During the crossing of the cold cloud, Earth would have been exposed to high HEP radiation for approximately 3 months per year while immersed in the heliosheath and increased GCR radiation for the rest of its trajectory. The duration of the crossing will depend largely on the as-yet-unknown size of these clouds. Only the central cloud of the nearby Local Ribbon of Cold Clouds (LRCC)[56], the Local Leo of Cold Clouds was studied[32], and not the broader LxCC environment (Figure 1b). It is reasonable to envision the morphology as sheet-like, with sizes that vary between aus[32] to parsecs. Recent observations argue that the clouds in LRCC have to be small[57]. Based on recent observations of supernova remnants it is reasonable to assume that the clouds could be filamentary[58]. Hence, the duration could be between yr to $10^5$ yrs considering the relative speed of 14km/s=14pc/Myr.

A prolonged ($10^3$- $10^5$ yr) event that increases the GCR flux or the HEP radiation to Earth should be detectable in cosmogenic isotopes signatures preserved in the geologic record. More recent radiation events are detected in cosmogenic $^{14}C$ in tree rings and $^{10}Be$ in ice cores[59,**87-89**]. While the half-life of $^{14}C$ is too short to record events occurring more than ~50,000 years ago, the half-life of $^{10}Be$ is sufficiently long to potentially capture any large radiation changes within the past 10 million years. However, ice core records only go back to ~1 million years ago, so one must turn to marine sediments or other archives to explore intervals further back in time.

Typically, marine sediment records that span multiple millions of years accumulate slowly (~cm/kyr) so they are best suited to capture events lasting 1000 years or longer. This is because bioturbation by organisms living within the sediment tend to smooth the record over time. Shorter events may be difficult to resolve in such slowly accumulating records. Ferromanganese crusts growing on the seafloor provide their own archival challenges, as their even slower rates of growth (~several mm/Myr) mean that each mm of material captures the integrated signal of ~$10^5$-$10^6$ years of time (e.g. ref. 14).

However, there are additional factors that impact $^{10}$Be concentrations in marine sediments that are independent of the cosmic ray flux. These factors include changes in Earth's magnetic field strength, dilution by other sediment components, and changes in sediment accumulation rate[60]. Therefore, simply identifying a change in $^{10}$Be concentration or $^{10}$Be/$^9$Be ratios (or not) does not rule out the possibility that changes in $^{10}$Be production did (or did not) occur. The largest signatures in $^{10}$Be are due to HEPs and are dependent on the size of the cloud (and duration of the crossing)[84].

Our current paper focuses on the likely radiation event between 2-3 million years ago. However, we expect another radiation increase between 6-7 million years ago, during the crossing of the edge of the Local Bubble[11]. Interestingly recently, ref. 61 reported a prolonged cosmogenic $^{10}$Be anomaly at 10 Myr. There have been recent studies reporting no particular peak in $^{10}$Be in the past 2-3 million years[62] while others show much more structure, in particular across the 2-3 million year interval[63]. This could be because some of the $^{10}$Be records are not ideal for detecting a radiation change in $^{10}$Be accumulation rates because the archives examined are dated assuming a constant input of $^{10}$Be[62,64]. Such studies should be revisited with careful high-resolution analysis and any record of $^{10}$Be accumulation should be independently dated. If indeed the $^{10}$Be records that show no variation hold[62] that will put a limit on the size of the cloud that the Sun had encountered.

What is the probability of crossing cold clouds? Cold clouds are very rare in the interstellar medium (less than 1% by volume of the interstellar medium) – they also might evolve. Under the assumption that a) the cold dense clouds in the Local Ribbon of Cold Clouds (LRCC) didn't evolve in the last 2-3 million years and b) that the Local Ribbon of Cold Clouds move as one solid object (with one speed). Ref. 10 derived the three components of velocity between the Sun and the LRCC and found that the Sun had 68% probability (1σ) of crossing the tail end of the LRCC in the last 2-3 million years. However, because these clouds are so small, to envision that we crossed exactly one of these clouds is unlikely (e.g., area of 1σ covers is tiny – only 1.3% of the sky (576 square degrees)).

In the case of the encounter of the Local Bubble, ref 11 estimated the cross section of the Sun encountering a cloud with a typical density of ~283 cm$^{-3}$ as it passes through superbubbles like the Local Bubble. They found that Clouds with density of 283cm$^{-3}$ cover a small but nonzero (4.6%) fraction of the surface of the Local Bubble, making an encounter plausible.

Even though these probabilities are low, they are not negligible. The $^{60}$Fe data tells us that something happened 2-3 Mya and 6-7 Mya that is consistent with the collapse of the heliosphere.

One interesting aspect is that the increased radiation of HEPs is unique to the encounter of the heliosphere with cold clouds. The supernova scenario predicts a contraction of the heliosphere but not to sub-au scales. In that case the Termination Shock will be at large distances > 10 au[4] and the radiation will be substantially weaker at Earth.

This radiation is the one outside of Earth's magnetosphere. Earth's magnetic field and its atmosphere affect the amount of radiation that reaches Earth's surface. In order to estimate the ionization on Earth's surface, one needs to consider the interaction of the radiation on ionizing the atmosphere with tools like Monte Carlo CORSIKA models[65]. Future aspects to consider include how Earth magnetosphere will be modified as Earth dips in and out of the compressed heliosphere and how its magnetic field modulates the energetic particles. The effect of the shielding for periods when Earth was outside the heliosphere is briefly discussed in ref. 66 and additional exploration is left for future work.

Of course, Earth's magnetic field protects the atmosphere and surface to some extent from incoming radiation (at least in latitudes that are not polar). Due to complex circulation in Earth's geodynamo, the dipole strength of its magnetic field, and consequently the protection its magnetic field provides from cosmic radiation, varies over time[67-68]. When the Earth's magnetic field reverses polarity, as has occurred at irregular intervals many times in the geologic past[69], the geomagnetic field can decrease by nearly an order of magnitude[70]. It is not known how long this process takes, though it is generally thought to be a few thousand years[68]. It is interesting to note that there was such a reversal, the Gauss-Matuyama reversal, at 2.58 Ma, around the time of our predicted crossing of the cold cloud. If the crossing occurred coincidentally with this reversal, then the consequent increase of radiation will easily reach the top of Earth's atmosphere.

It is interesting to comment on the effect of HEPs for $^{60}$Fe. The flux of cosmogenic $^{60}$Fe produced by cosmic rays in the atmosphere was estimated by ref. 71 to be three orders of magnitude lower than the pulse $^{60}$Fe flux 2-3 Myr ago based on deep-sea crust measurements. The most prominent pathway for producing 60Fe in the atmosphere from GCR protons is by spallation of krypton, which reaches a peak cross section at 800 MeV —this energy is where the HEP spectrum (Fig. 1) is below the present-day cosmic ray spectrum. Therefore, we do not expect HEPs to generate any significant amount of $^{60}$Fe. As commented in the introduction, the increase in flux of $^{60}$Fe will come in form of dust that gets incorporated into massive clouds existent in the interstellar medium. As Earth goes through a massive cloud enriched with $^{60}$Fe, the flux of terrestrial $^{60}$Fe will increase.

The increased radiation could have effects on both climate and life on Earth's surface. There have been works[72] that explore the possible effect of increased radiation at TeV GCR energies and above from supernova explosions on biodiversity dynamics, for example attempting to link them to mass extinctions. Here we are interested in radiation at a much lower energy range. Cosmic rays at GeV energies and HEPs interact with the Earth's climate by precipitating into the atmosphere and ionizing molecules such as nitrogen and oxygen gas, which then form NOx and HOx compounds. These compounds react to catalytically destroy stratospheric ozone, reducing stratospheric temperatures. Ozone depletion is associated with a cooling of the polar vortex, which can produce regional increases in surface temperature[73-74]. Other works suggest that increased geomagnetic activity may drive increased surface temperature variability due to this effect[75]. As discussed above, the potential climate effects due to increased hydrogen during the exposure of Earth to the ISM may also be important and are being revisited.

Earth's history has been marked by a changing climate, and such changes have far-reaching implications for geological, biogeochemical, and evolutionary processes. For example, there were rapid global cooling events 13-14 Mya, 6-7 Mya, and 2-3 Mya, and a general increase in variability over the past 10 Mya, as indicated by oxygen isotopes ($^{18}$O and $^{16}$O) measured in the skeletons of microscopic foraminifera found in deep sea sediment cores[20-21]. These events led to major reorganizations of the world's biomes and changes to diversification patterns of different groups of organisms. For example, global cooling between 2-3 Mya is associated with an increased presence of Northern Hemisphere ice sheets[20]. However, the forcing mechanisms behind such historical climate shifts, such as sudden cooling events, remain poorly understood and represent a major open question in science. Small perturbations to Earth's radiative balance, driven by changes in solar radiation or atmospheric composition, for example, can be amplified by internal feedbacks within the climate system and result in much larger and more prolonged impacts on global climate than the original forcing alone[76]. Studies of astronomical effects on Earth's past climate typically only consider how changes in Earth's orbital geometry affect incoming solar radiation (e.g., the Milankovitch cycles)[20,76].

Further studies of past climate should consider the impact of the heliosphere itself and include not only the impact of increased neutral density, but of increased and varied radiation as is considered in modern climate models[22,78]

In principle, our proposed radiation changes could also impact organismal mutation rates, aging, speciation, and extinction rates, and thus broad patterns of diversification. When cosmic radiation collides with the upper atmosphere, it showers particles down toward the surface of the Earth. While most of the particles are filtered out before reaching the surface, neutrons and muons are not, and the latter can penetrate deep into the ocean and deep underground, encountering nearly every organism on Earth. These types of ionizing radiation can impact biological tissue and induce damage, increase the production of free radicals, which in turn can lead to DNA damage and mutations[79-80]. Increased germline mutation rates could have had effects on the evolution and diversification of life, while increased somatic mutation rate could lead to an increase in cancer rates[80] or changes that mimic aging[81], and other potentially negative effects on organisms. However, there is not a good understanding of how different levels of cosmic radiation affect biological organisms, and whether the magnitude of changes predicted here would have affected the evolution of life. Thus, any potential effects of heliospheric changes on Earth's life via variation in incident radiation remain speculative and should be a topic for future work.

We hope that this paper will inspire further exploration of this new paradigm that the location of the Sun through its trajectory had consequences on Earth. With the advent of the Gaia mission there has been a revolution in our ability to locate precisely which structures the Sun would have encountered on its path. Thanks to new advances[82], we can now map out the Sun's interstellar environment out to distances of 4000 light-years, which corresponds to being able to trace the Sun's trajectory back roughly 60 Mya. The 60-million-year timescale is also the timescale over which we have global climate records constructed from deep-sea sediments. We are excited to continue this exploration in future works.

**Acknowledgments:** This work is supported by NASA grant 80NSSC22M0164, 18-DRIVE18_2-0029 as part of the NASA/DRIVE program titled "Our Heliospheric Shield". For more information about this center please visit: https://shielddrivecenter.com/. EPE was supported by JSPS KAKENHI (24K01785).

**Funding:** This work is supported by NASA grant 80NSSC22M0164, 18-DRIVE18_2-0029 as part of the NASA/DRIVE program titled "Our Heliospheric Shield". For more information about this center please visit: https://shielddrivecenter.com/. EPE was supported by JSPS KAKENHI (24K01785).

**Author contributions:** Conceptualization: M.O., E.P.E. and A.L; Methodology: M. O. MHD code results; J.G. and A.C. estimation of the radiation; Interpretation and Conclusions, Writing: All authors were involved

**Data and materials availability:** Original data created for the study are of will be available in a persistent repository upon publication in https://zenodo.org/

**Figures and Tables**

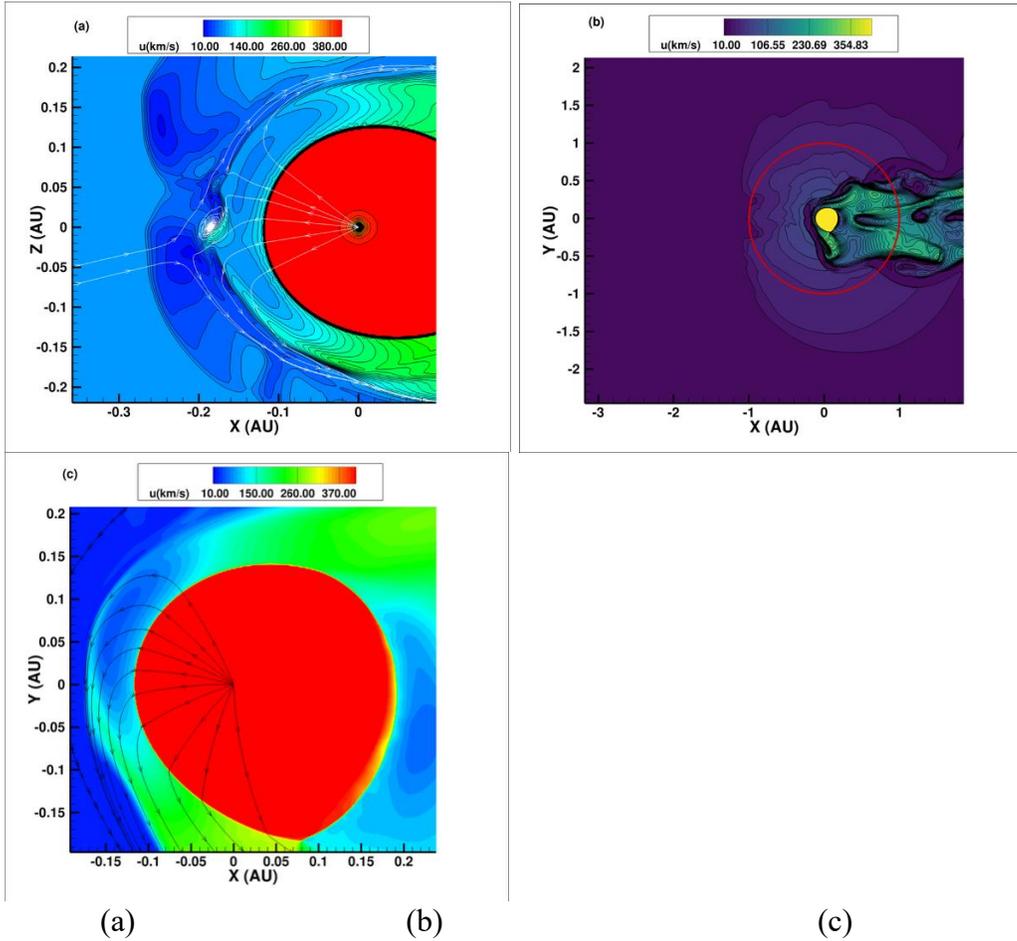

(a) (b) (c)

**Figure 1. Heliosphere 2-3 Myr ago** –speed. (a) meridional plane (b) equatorial plane showing the trajectory of Earth. When Earth will be inside the heliosphere the Galactic Cosmic Rays will be filtered by the heliosphere and there will be an increase of HEPs. (c) zoom of panel b on the termination shock. The termination shock at these small distances is a parallel shock.

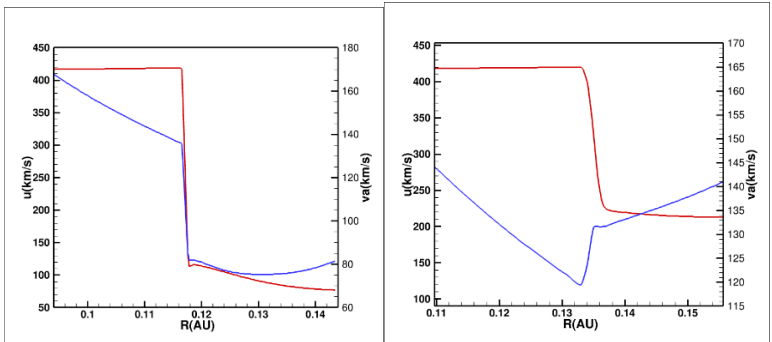

(a)                                             (b)

**Figure 2. Termination Shock 2-3 Myr ago (a) nose; (b) flanks showing the solar wind speed (red) and alfven speed (blue)** The shock at the nose is strong with compression ratio 3.62 and at the flanks with 1.88.

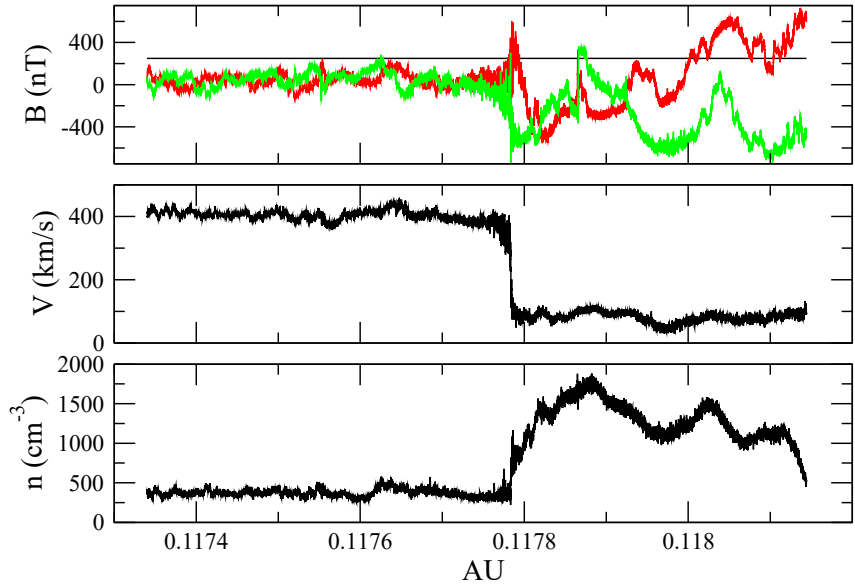

(a)

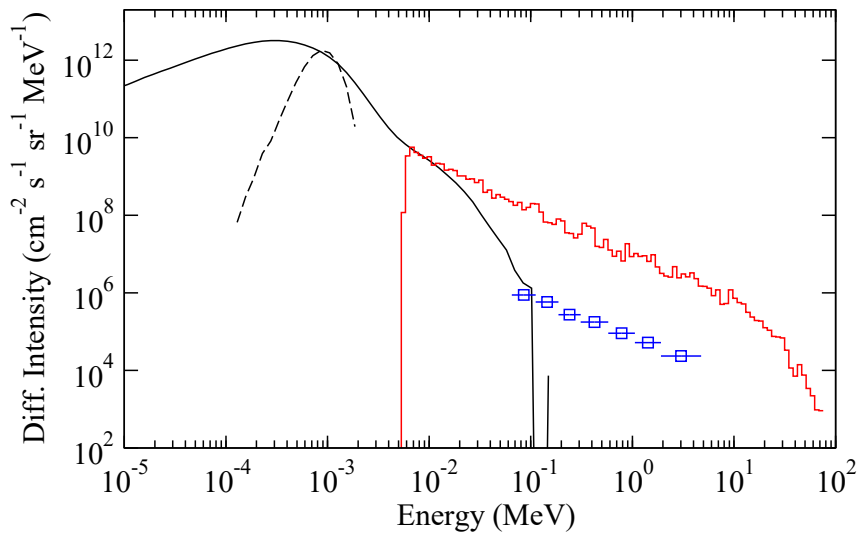

(b)

**Figure 3. Radiation spectrum outside Earth magnetosphere when Earth is inside the heliosphere** (a) Magnetic field, solar wind velocity and density in the near vicinity of the TS, along the nose direction. Shown are the components of the magnetic field. Black is the x component (constant because this is a 1D simulation and div B = 0), y is red, and z is green. (b) Radiation spectrum outside Earth magnetosphere when Earth is inside the heliosphere: Simulated differential intensity of protons at the termination shock located at about 0.118 au as a black curve and red histogram, along with data from the ACE/EPAM instrument for a large solar-energetic particle event that occurred on DOY 302 2003. The data was taken immediately after that passage of a large CME-driven shock associated with this event. It is the same data as that presented in Figure 2 of ref (83). This is among one of the largest SEP events observed. The black curve is from a hybrid simulation of a quasi-parallel collisionless shock using parameters consistent with the solar wind and interplanetary magnetic field at 0.118 au. The red histogram is the result from a 1D-spherical calculation of particle acceleration at a spherical shock located at this distance. It is based on a model which is a solution to the Parker transport equation. See text for details on the two models used. We name these energetic particles *heliospheric energetic particles* or HEPs.

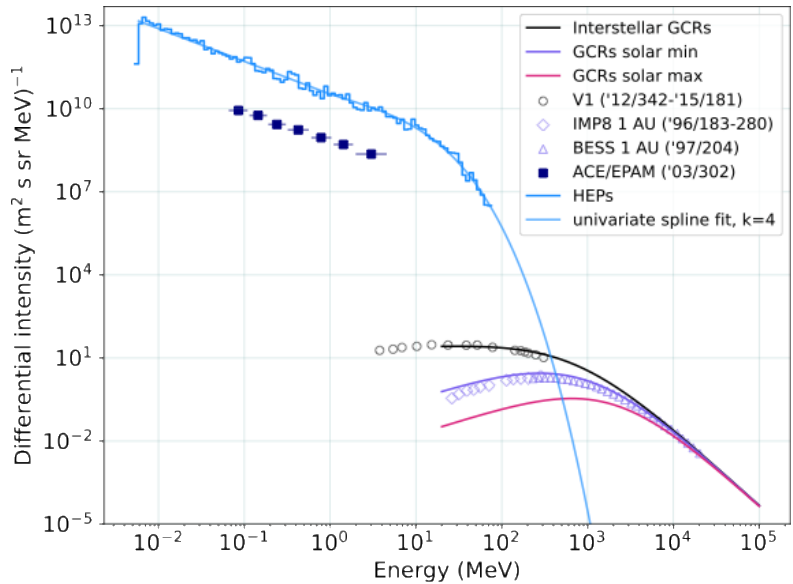

**Figure 4. Radiation exposed outside of Earth magnetic field.** Simulated proton differential energy spectra of the heliospheric energetic particle (HEP) flux at the termination shock of the compressed heliosphere and applied fit function (blue lines), unmodulated interstellar flux of GCRs (black line), GCRs modulated for typical solar minimum conditions (purple line), and typical solar maximum conditions (pink line). Navy marks represent the energetic particle flux detected by the ACE/EPAM instrument during the solar energetic particle storm of DOY 302 2003 (the so-called Halloween 2003 event). Black marks represent cosmic ray detections from Voyager 1 from DOY 342 of 2012 to DOY 181 of 2015, and purple marks from a BESS balloon flight in 1997 and from IMP8 in 1996 at 1 AU. These data are shown as in ref. 1, for hydrogen. The low-MeV HEP flux is ~100 times higher than the solar proton storm of 2003, and ~9 orders of magnitude greater than the interstellar GCR flux. The HEP spectrum is assumed to reduce to ~0 at energies > 1 GeV.